\documentstyle[aps,prl,twocolumn]{revtex}
\def\be{\begin{equation}}
\def\ee{\end{equation}}
\def\bea{\begin{eqnarray}}
\def\eea{\end{eqnarray}}
\def\bma{\begin{mathletters}}
\def\ema{\end{mathletters}}
\newcommand{\bra}[1]{\mbox{$\langle #1 |$}}
\newcommand{\ket}[1]{\mbox{$| #1 \rangle$}}

\newcommand{\proj}[1]{\ket{#1}\!\bra{#1}}

\begin{document}

\draft

\title{Entanglement of pure states for a single copy}

\author{Guifr\'e Vidal}

\address{e-mail: guifre@ecm.ub.es\\
Departament d'Estructura i Constituents de la
Mat\`eria\\ 
Universitat de Barcelona
\\ Diagonal 647, E-08028 Barcelona, Spain}

\date{\today}

\maketitle

\begin{abstract}
An optimal local conversion strategy between any two pure states of a bipartite system is presented. It is optimal in that the probability of success is the largest achievable if the parties which share the system, and which can communicate classically, are only allowed to act locally on it. The study of optimal local conversions sheds some light on the entanglement of a single copy of a pure state. We propose a quantification of such an entanglement by means of a finite minimal set of new measures from which the optimal probability of conversion follows.  
\end{abstract}


\bigskip 

A proper quantification of entanglement is a priority in quantum information theory, for many of its applications \cite{PW} rely on quantum correlations as a necessary resource. Such a quantification of the non-local resources of a state should provide us with a detailed account of which tasks can be accomplished with it, or more specifically -- since we are in the quantum kingdom --, with the probability with which a given task can be accomplished. 

 Our work addresses the quantification of the entanglement of pure states shared by two parties. So far its complete quantification has been achieved only in a very specific limit, namely when the parties share infinitely many identical copies of a given state \cite{BeDi,BeBe}. On the other hand some authors  (\cite{LoPop,Vid,Nie,BoVeKn}) have initiated the study of the entanglement of a finite number of copies of pure states. This finite framework, which is the one involved in the realistic situations one encounters in a lab, will be also the objective of this work. The approach taken here relies on the study of optimal local transformations and of magnitudes that have a monotone behavior under any local manipulation of the system, which will be referred to as {\em entanglement monotones}\cite{Vid}. We present a new set of such entanglement monotones and argue that they together quantify uniquely the entanglement of pure states in a physically relevant fashion. 

 Our starting question is the following: suppose that Alice and Bob share a pure entangled state $\Psi$ and that they would like to convert it into another pure entangled state $\Phi$. Which is the greatest probability of success in such a conversion if the two parties, which are classically communicated, are only allowed to act on the system locally? 

 We present here the answer to this question (see eq. (\ref{main})) \cite{comment}, together with an explicit local strategy achieving the optimal probability. We also investigate, and refute, a possible ordering on the set of pure states induced by such probability.  Finally some considerations regarding the nature of entanglement are made, and reversibility of optimal conversions, additivity of entanglement and uniqueness of their measure are argued not to hold.

 Let us start by considering the most general pure state of a bipartite system $\Psi \in{\cal C}^n\otimes{\cal C}^n$ and its Schmidt decomposition
\be
\ket{\Psi} = \sum_{i=1}^n \sqrt{\alpha_i} \ket{i_A~ i_B},~~~~\alpha_i\geq \alpha_{i+1}\geq 0,~\sum_{i=1}^n \alpha_i = 1,
\label{psi}
\ee
where  $\{\sqrt{\alpha_i}\}$ are the Schmidt coefficients of $\Psi$ and $\ket{i_A~ i_B}$ stands for $\ket{i_A}\otimes\ket{i_B}$, $\{\ket{i_A}\}_{i=1}^n$ and $\{\ket{i_B}\}_{i=1}^n$ being two local orthonormal bases depending on $\Psi$. Since Alice and Bob are allowed to perform local unitary transformations, and these are locally reversible, any two states with the same Schmidt coefficients are locally equivalent. Thus only the Schmidt coefficients are relevant as far as non-locality is concerned, and we may study, without loss of generality, the optimal local conversion of $\Psi$ into $\Phi$ satisfying 
\be
\ket{\Phi} = \sum_{i=1}^n \sqrt{\beta_i} \ket{i_A~i_B},~~~~\beta_i\geq \beta_{i+1}\geq 0,~\sum_{i=1}^n \beta_i = 1.
\label{phi}
\ee

{\bf Theorem:} Let us call $P(\Psi \rightarrow \Phi)$ the maximal probability of obtaining the state $\Phi$ from $\Psi$ by means of any local strategy. Then, in terms of the Schmidt coefficients of $\Psi$ and $\Phi$, we have
\be
P(\Psi \rightarrow \Phi) = \min_{l\in[1,n]} \frac{\sum_{i=l}^n\alpha_i}{\sum_{i=l}^n\beta_i}.
\label{main}
\ee
\vspace{3mm}

 Before proving this result, let us note here that for $\ket{\Phi} = \sum_{i=1}^m \frac{1}{\sqrt{m}}\ket{i_A~i_B}$ we recover the results obtained by Lo and Popescu in \cite{LoPop}, while the entanglement monotone $E_{k=2}$ reduces, in the two-qubit case (that is, $n=2$), to the {\it entanglement of single pair purification} introduced by Bose, Vedral and Knight in \cite{BoVeKn}.
\vspace{2mm}

{\bf Proof:} Optimality of eq. (\ref{main}) will be proved by
\begin{itemize}
\item
showing an explicit local strategy which converts $\Psi$ into $\Phi$ successfully with such probability, and by
\item introducing a family of entanglement monotones, denoted by $E_k(\rho)$, $k=1,\cdots,n$, and defined over the set of pure states as
\be
E_k(\Psi) \equiv \sum_{i=k}^n \alpha_i,
\label{pure}
\ee
whose monotonicity sets the upper bound 
\be
P(\Psi \rightarrow \Phi) \leq \min_{l\in[1,n]} \frac{\sum_{i=l}^n\alpha_i}{\sum_{i=l}^n\beta_i} =\min_{l\in[1,n]} \frac{E_l(\Psi)}{E_l(\Phi)}.
\label{upperbound}
\ee
\end{itemize}

Indeed, suppose  that there is a local strategy with probability of success $P'$ greater than this upper bound, and that the minimum in eq. (\ref{upperbound}) is for $l = l_{\star}$. Before the conversion the amount of the monotone $E_{l_{\star}}$ is $E_{l_{\star}}(\Psi)$, and after the conversion it would be, on average, at least --since we may be neglecting positive contributions coming from unsuccessful conversions-- $P'E_{l_{\star}}(\Phi) > E_{l_{\star}}(\Psi)$, which would mean an increase of this (non-increasing) entanglement monotone, and would lead therefore to a contradiction.

That eq. (\ref{pure}), together with the convex roof extension of $E_k$ to mixed states
\be
E_k(\rho) \equiv \min_{\Upsilon_{\rho}} \sum_j p_j E_k(\psi_j),
\label{mixed}
\ee
(here the minimization is to be performed over all the pure-state ensembles $\Upsilon_{\rho}=\{p_j,~\psi_j\}$ realizing $\rho$, i.e. such that $\rho=\sum_j p_j\proj{\psi_j}$), defines an entanglement monotone for each $k$ follows from the fact that $E_k(\Psi)$ can be written as
\be
E_k(\Psi) = f_k(\mbox{Tr}_A \proj{\Psi}),
\ee
where $f_k(\sigma)\equiv \sum_{i=k}^n \alpha_i$ --that is the sum of the $n-k+1$ smallest eigenvalues of $\sigma$-- is a unitarily-invariant, concave function of $\sigma$, and from Theorem 2 in \cite{Vid}. That $f_k(\sigma)$ is a concave function follows from the "Ky Fan's Maximum Principle" \cite{Bha}.
Therefore what remains to be shown is that there is a local conversion strategy compatible with eq. (\ref{main}).

 Notice, first, that eq. (\ref{upperbound}) implies that if the number of non-zero Schmidt coefficients of $\Psi$ is smaller than that of $\Phi$, then $P(\Psi\rightarrow \Phi) = 0$, as it is already well known \cite{LoPop}.  Therefore we will assume from now on that $\Psi$ has at least as many non-vanishing Schmidt coefficients as $\Phi$. We will also assume, for simplicity sake, that $\alpha_n > 0$ (by lowering de dimension $n$ of the original local Hilbert spaces if needed). 

 The optimal local conversion strategy we present here consists of two steps. In the first one the parties convert, with certainty, the initial state $\Psi$ into a temporary pure state $\Omega$, by making use of a local strategy recently proposed by Nielsen \cite{Nie}. In a second step $\Omega$ is converted into $\Phi$ by means of a local measurement, $P(\Psi\rightarrow \Phi)$ being the probability that this last conversion be successful.

 Let us thus call $l_1$ the smallest integer $\in [1,n]$ such that
\be
\frac{\sum_{i=l_1}^n\alpha_i}{\sum_{i=l_1}^n\beta_i} = \min_{l\in[1,n]} \frac{\sum_{i=l}^n\alpha_i}{\sum_{i=l}^n\beta_i} \equiv ~r_1~~ (\leq 1).
\ee
It may happen that $l_1 = r_1 = 1$. If not, it follows from the equivalence
\be
\frac{a}{b} < \frac{a+c}{b+d} \Leftrightarrow \frac{a}{b} < \frac{c}{d}~~~~~(a,b,c,d > 0)
\ee
that for any integer $k \in [1, l_1-1]$
\be
\frac{\sum_{i=k}^{l_1-1}\alpha_i}{\sum_{i=k}^{l_1-1}\beta_i} > r_1.
\ee
 Let us then define $l_2$ as the smallest integer $\in [1,l_1-1]$ such that
\be
r_2\equiv \frac{\sum_{i=l_2}^{l_1-1}\alpha_i}{\sum_{i=l_2}^{l_1-1}\beta_i} = \min_{l\in[1,l_1-1]} \frac{\sum_{i=l}^{l_1-1}\alpha_i}{\sum_{i=l}^{l_1-1}\beta_i} ~~~ (> r_1).
\ee
 Repeating this process until $l_k=1$ for some $k$, we obtain a series of $k+1$ integers $l_0 > l_1 > l_2 > \cdots >l_k$ ($l_0 \equiv n+1$) and $k$ positive real numbers $0 < r_1 < r_2 < ... < r_k$, by means of which we define our temporary (normalized) state $\ket{\Omega} = \sum_{i=1}^n \sqrt{\gamma_i} \ket{i_A~i_B}$, where

\be
\gamma_i \equiv r_j\beta_i  ~~~\mbox{ if } i \in [l_j, l_{j-1}-1],~~~\mbox{i.e.,}
\ee

\be
\vec{\gamma} = \left[ \begin{array}{c} 
r_k\left[ \begin{array}{c} 
   \beta_{l_k} \\ \vdots \\ \beta_{l_{k-1}-1}
   \end{array} \right] \\
\vdots \\
r_2\left[ \begin{array}{c} 
   \beta_{l_2} \\ \vdots \\ \beta_{l_1-1}
   \end{array} \right]\\
r_1\left[ \begin{array}{c} 
   \beta_{l_1} \\ \vdots \\ \beta_{l_0-1}
   \end{array} \right]
\end{array} \right].
\ee
Notice that by construction
\be
\sum_{i=k}^n \alpha_i \geq \sum_{i=k}^n \gamma_i~~~~\forall k\in[1,n]
\ee
(or, equivalently, $\sum_{i=1}^k \alpha_i \leq \sum_{i=1}^k \gamma_i~~~~\forall k\in[1,n]$, that is to say, $\vec{\alpha}$ is majorized by $\vec{\gamma}$). Consequently Nielsen's local strategy shown in \cite{Nie} can be applied in order for the parties to obtain the state $\Omega$ from $\Psi$ with certainty.

 Let us consider now the positive operator $\hat{M}:{\cal C}^n\longrightarrow {\cal C}^n$
\be
\hat{M}\equiv \left[ \begin{array}{cccc} 
\hat{M}_k &   &   & \\
    & \ddots  &   & \\
    &   & \hat{M}_2  & \\
    &   &   & \hat{M}_1\\
\end{array} \right] = \hat{M}^{\dagger},
\ee
where
\be
\hat{M}_j \equiv \sqrt{\frac{r_1}{r_j}} \hat{I}_{[l_{j-1}-l_{j}]}~~~j=1,\cdots,k,
\ee
is proportional to the identity in a $(l_{j-1}-l_{j})$-dimensional subspace of ${\cal C}^n$. It satisfies that $0\leq \hat{M} \leq I$, so that together with $\hat{N}\equiv \sqrt{1-\hat{M}^2}$ it defines a generalized measurement of two outcomes ($\hat{M},\hat{N} \geq 0; \hat{M}^{\dagger}\hat{M}+\hat{N}^{\dagger}\hat{N} = \hat{I}$) that Alice (for instance) can perform locally. Since $\hat{M}\otimes \hat{I}_B \ket{\Omega} = \sqrt{r_1}\ket{\Phi}$, the whole local strategy allows to obtain the pure state $\Phi$ from $\Psi$ with (optimal) probability $P(\Psi \rightarrow \Phi) = r_1$. Notice that $\hat{N}\otimes \hat{I}_B \ket{\Omega}$ is an unnormalized, pure (often entangled) state with less non-vanishing coefficients than $\Phi$, so that, as expected, one can not use it to obtain $\Phi$. This ends the proof of eq. (\ref{main}).$\Box$

\vspace{3mm}

 Notice that this strategy can be minimally implemented, for instance, with local measurements on Alice's side, one way classical communication (from Alice to Bob) and local unitary transformations on both sides, these three types of allowed operations being performed several times (the number of operations will depend on the two states, but is of order $n$). Notice also that this strategy is not the simplest optimal one since optimal local conversion strategies must exist involving only one measurement on Alice side, plus one transmission of classical bits from Alice to Bob, plus one locally unitary transformation on each side (see \cite{LoPop}). 

 Let us briefly consider an alternative scenario where eq. (\ref{main}) can be applied. Suppose that, as before, the parties start sharing the pure state $\Psi$, but that now their aim is to obtain (on average) the greatest number of copies of the state $\Phi$, say $m_{\Psi\rightarrow\Phi}^{MAX}$. In this case the optimal strategy involve, if possible, local conversions into several copies of $\Phi$, and this is not ruled in general by eq. (\ref{main}). However, there are circumstances in which $m_{\Psi\rightarrow\Phi}^{MAX}=P(\Psi \rightarrow \Phi)$. Indeed, let $n_{\psi}$ denote the number of non-vanishing Schmidt coefficients of the entangled state $\psi$, and recall that $n_{\psi^{\otimes N}} = n_{\psi}^N$. Then,
\be
n_{\Psi} < n_{\Phi}^2 \Rightarrow P(\Psi \rightarrow \Phi^{\otimes N})=0~~ \forall~~ N \geq 2
\ee
implies that the greatest number $m_{\Psi\rightarrow\Phi}^{MAX}$ of copies of $\Phi$ the parties can obtain locally from $\Psi$ is also given by $P(\Psi \rightarrow \Phi)$ when $n_{\Psi} < n_{\Phi}^2$.

 Let us move to consider now the following question: Is there any order in the space of entangled pure states that can be derived from eq. (\ref{main})? In \cite{Nie} a partial order on the entangled pure states was obtained according to whether, given two states $\Psi_1$ and $\Psi_2$, one of them can be converted locally into the other with certainty, say $\Psi_1$ into $\Psi_2$. If so, $\Psi_1$ can be said to contain at least as much entanglement as $\Psi_2$, in the sense that any non-local resource that $\Psi_2$ may contain, it is automatically contained, in at least the same amount, also in $\Psi_1$, and, again, the non-local resources needed to obtain $\Psi_1$ suffice to create $\Psi_2$. But if on the contrary $\Psi_1$ and $\Psi_2$ are such that none of them can be converted into the other with certainty, then their entanglement is incommensurable according to this criteria. One might be tempted to extend such a partial order to the whole set of pure states by saying that the state $\Psi_1$ is more entangled than $\Psi_2$ if, and only if, $P(\Psi_1 \rightarrow \Psi_2)~>~P(\Psi_2 \rightarrow \Psi_1)$. However, the following example shows that this order would be ill-defined: consider three states $\Psi_k \in {\cal C}^4\otimes{\cal C}^4$, the square of the Schmidt coefficients of the $k$-th state being $\vec{\alpha}_{k}$, where
\begin{eqnarray}
\vec{\alpha}_{k=1} &\equiv& \frac{1}{144}( 108, 12, 12, 12),\nonumber\\
\vec{\alpha}_{k=2} &\equiv& \frac{1}{144}( 66,  66,  6, 6),\nonumber\\
\vec{\alpha}_{k=3} &\equiv& \frac{1}{144}( 47, 47, 47, 3).
\end{eqnarray}
Then such an ordering relation leads to the following contradiction:
\be
\Psi_1 < \Psi_2 < \Psi_3 < \Psi_1.
\ee

 Finally, we would like to analyze what conclusions can be drawn from eq. (\ref{main}) regarding the quantification of entanglement of a shared state, understood in relation to the non-local resources that characterize it. One can consider, e.g., both {\em how many} such non-local resources are needed to create the state and {\em how many} of them can be extracted from it, in terms of other shared states.

For pure states of a bipartite system the entropy of entanglement $E(\Psi^{\otimes N})$ \cite{BeDi}, and therefore one sole --and unique \cite{PopRo}-- parameter, quantifies asymptotically the non-local resources of a huge number $N$ of copies of a given shared state $\Psi$. It turns out that in such a context optimal local conversions are reversible, and that entanglement behaves as an additive property of the quantum world.

 We have considered in the present work the optimal local conversion of single copies of pure states, which falls far from the large-N asymptotic case. This finite scenario is relevant in the light of the state of present technology, for it is not clear yet how to perform certain local transformations in the space of a large number of copies which are a necessary ingredient in the asymptotic conversions so far exposed \cite{BeBe}. But even if one knows how such local transformations can be performed in a lab, the finite scenario is important on its own, since it describes the local resources involved in any local manipulation of a finite number of copies of pure states. Eq. (\ref{main}) teaches us the following qualitative facts about entanglement:

\begin{enumerate}
\item Irreversibility: the optimal local conversion between any two states with non-identical Schmidt coefficients is always an irreversible process. Here irreversible means that the parties can not, with certainty, convert locally one state into another and then get the initial state back. This general result, which was proved in \cite{Vid} and follows also from \cite{Nie}, does not hold asymptotically. 

\item More than one measure: the quantification of entanglement, in the sense exposed above, requires more than just one measure \cite{Vid}. For pure states in ${\cal C}^n\otimes{\cal C}^n$, the $n-1$ entanglement monotones $E_k~(k=2,\cdots,n)$ are a minimal set of non-increasing parameters providing a detailed and straightforward account of their non-local resources. They can be regarded as the measures of the entanglement of pure states in a similar sense as the entropy of entanglement is their measure in the asymptotic limit.

\item Non-additivity: the non-local resources of entangled states are not additive in general, in that, for instance, two parties can often extract more such resources from two copies of a given shared state $\Psi$, i.e. from $\Psi\otimes\Psi$, than twice what they can obtain from one single copy $\Psi$ \cite{example}. From this point of view it is artificial to take additivity as an a priori requirement for any good measure of entanglement\cite{Vid}. Thus additivity of the entanglement of pure states in the asymptotic limit is a remarkable result, rather than an a priori constraint, which follows from the additivity of the entropy of entanglement and from the existence of reversible asymptotic conversions (as the ones in \cite{BeBe}).
\end{enumerate}

 Summarizing, given two pure shared states $\Psi$ and $\Phi$, the highest probability $P(\Psi \rightarrow \Phi)$ of success in the conversion of $\Psi$ into $\Phi$ by means of any local strategy can be used to quantify their entanglement. We have shown that for pure states of a bipartite system the entanglement monotones $E_k$ provide $P(\Psi \rightarrow \Phi)$, since this probability is the greatest one compatible with the monotonicity of $E_k$. The explicit expression for $P(\Psi \rightarrow \Phi)$ shows that the entanglement of a pure state $\Psi$ behaves essentially differently from that of $\Psi^{\otimes N}$ for very large $N$.  
 
 There are many open problems regarding finite entanglement. It would be interesting to derive equivalent results for pure states shared by three or more parties. Also, to extend the results presented here to mixed states \cite{mixed}. A way to proceed is by studying concrete local conversion strategies, which mean lower bounds on the optimal probability of success in the conversion, and by identifying new entanglement monotones, since each one implies an upper bound. In this scheme it becomes reasonable to demand, as an a priori requirement, only monotonicity under local manipulations in order for a magnitude to be a candidate for a measure of entanglement.

 The author is most grateful to Rolf Tarrach for his thorough reading of the manuscript, comments and suggestions. Comments are also acknowledged to Maciej Lewenstein. The author thanks Maciej Lewenstein, Anna Sanpera and Christel Franko for their hospitality in Hannover. Financial support from CIRYT, contract AEN98-0431 and CIRIT, contract 1998SGR-00026 and a CIRIT grant 1997FI-00068 PG are also acknowledged.


\begin{references}
\bibitem{comment} Notice that in \cite{LoPop} the same question was already answered when the final state is of the form $\ket{\Phi} = \sum_{i=1}^m \frac{1}{\sqrt{m}}\ket{i_A~i_B}$ for some $m$.
\bibitem{example} Take, as a concrete example, $\ket{\Psi}=\frac{1}{\sqrt{2}}(\ket{1_A~1_B}+\ket{2_A~2_B})$ and $\ket{\Phi} = \frac{1}{\sqrt{3}}(\ket{1_A~1_B}+\ket{2_A~2_B}+\ket{3_A~3_B})$. Then, for instance, $1 = P(\Psi\otimes\Psi \rightarrow \Phi) > 2 P(\Psi \rightarrow \Phi) = 0$.
\bibitem{mixed} In the particular case of a two-qubit system (${\cal C}^2\otimes{\cal C}^2$), a closed expression for the optimal probability of locally converting any pure state $\Psi$ into any mixed state $\rho$ has been obtained. It reads $P(\Psi \rightarrow \rho) = \min (\frac{E_2(\Psi)}{E_2(\rho)},1)$, where $E_2(\rho)$ is an explicit expression of $\rho$. This result, together with an optimal local strategy for such a conversion, will be presented somewhere else.
\end{references}
\end{document}